\documentclass[12pt,a4paper]{article}

\usepackage{colortbl}
\usepackage{ascmac}
\usepackage{bm}
\usepackage[dvipdfmx]{graphicx}
\usepackage{amsmath}
\usepackage{amsfonts}
\usepackage{multirow}

\def\tr{\mathop{\rm tr}\nolimits}
\def\Pexp{\mathop{\rm Pexp}\nolimits}
\def\diag{\mathop{\rm diag}\nolimits}

\newcommand{\wt}{\widetilde}

\newcommand{\RR}{\mathbb{R}}

\newcommand{\rap}[2]
{\setbox1=\hbox{#1}%
\setbox2=\hbox to\wd1{\hss #2\hss}%
\mbox{\rlap{\box1}\box2}}

\newcommand{\sla}[1]{\rap{$#1$}{$\backslash$}}
\newcommand{\sci}{\sharp}
\newcommand{\sch}{\natural}

\begin{document}


\begin{titlepage}
\title{
\vspace{-1.5cm}
\begin{flushright}
{\normalsize TIT/HEP-705\\ June 2025}
\end{flushright}
\vspace{1.5cm}
\LARGE{D1-brane correction to a line operator index}}
\author{
Yosuke {\scshape Imamura \footnote{E-mail: imamura@phys.sci.isct.ac.jp}} and
Akihiro {\scshape Sei \footnote{E-mail: a.sei@th.phys.titech.ac.jp}} \\
\\[18pt]
{\itshape Department of Physics, Institute of Science Tokyo}, \\ {\itshape Tokyo 152-8551, Japan}
}
\date{}
\maketitle
\thispagestyle{empty}
\begin{abstract}
Wilson line operators in the rank $k$ totally symmetric tensor representation of ${\cal N}=4$ $U(N)$ sypersymmetric Yang-Mills theories
are expected to be realized as D3-branes expanded in $AdS_5$.
Although
there is a mismatch
between the corresponding line operator indices even in the large $N$ and large $k$ limit,
it is possible to calculate the finite $k$ correction on the AdS side as the contribution from D1-branes.
We analyze D1-brane fluctuation modes and calculate the leading finite $k$ correction to the line operator index
on the AdS side.
\end{abstract}

\end{titlepage}

\tableofcontents
\section{Introduction}
Line operators are basic and important observables in gauge theories,
which can be used to detect phases of the system.
In the context of AdS/CFT correspondence \cite{Maldacena:1997re,Gubser:1998bc,Witten:1998qj},
the corresponding objects are branes probing
the corresponding geometric structure on the gravity side \cite{Witten:1998zw}.
In the recent progress concerning the detailed analysis of BPS operators
and the corresponding black hole systems
\cite{Hosseini:2017mds,Cabo-Bizet:2018ehj,Choi:2018hmj,Amariti:2019mgp,
Chang:2013fba,Chang:2022mjp,Choi:2022caq,Choi:2023znd,Kim:2023sig,Choi:2023vdm,Choi:2024xnv,deMelloKoch:2024pcs},
line operators
in different gauge group representations
may be useful
to probe objects residing in the AdS.

To perform such analyses, it is important to understand
the detailed relations between line operators in different representations
and the corresponding objects on the gravity side.
We consider AdS$_5$/CFT$_4$ correspondence for
the ${\cal N}=4$ $U(N)$ supersymmetric Yang-Mills theory.
For the fundamental representation, the corresponding object is
a fundamental string worldsheet with its end on the line on the AdS boundary \cite{Rey:1998ik,Maldacena:1998im}.
It is also known that lines in totally symmetric and totally anti-symmetric tensor representations
are realized by BPS configurations of D3 and D5-branes, respectively \cite{Drukker:2005kx,Yamaguchi:2006tq}.
The rank $k$ is determined by the fundamental string charge carried by the D-brane.
For circular Wilson loops in $S^4$ or $\RR^4$,
it was shown that the expectation values
are correctly reproduced with the D-branes \cite{Drukker:2005kx,Yamaguchi:2006tq,Hartnoll:2006is,Yamaguchi:2007ps}.
It was also shown in
\cite{Gomis:2006sb,Gomis:2006im}
that line operators in general representations
can be realized by multiple D3-branes and multiple D5-branes
in complementary ways,
and the corresponding classical supergravity configurations
were constructed \cite{Yamaguchi:2006te,Lunin:2006xr,DHoker:2007mci}.

Similar analyses were done for line operators in $S^3\times S^1$.
The partition function in $S^3\times S^1$ with supersymmetric boundary conditions
is the superconformal index \cite{Romelsberger:2005eg,Kinney:2005ej}.
With the line operator insertion,
it is called the line operator index \cite{Dimofte:2011py,Gang:2012yr,Drukker:2015spa}.
See also \cite{Hatsuda:2023iwi,Guo:2023mkn,Hatsuda:2023imp} for analytic formulas for line operator indices.
The line operator insertion partially breaks supersymmetry,
and to keep the system BPS, we need to imposes a restriction
on the values of the fugacities of the index,
and it requires us to take the
Schur limit.
The Schur limit of the superconformal index,
the Schur index \cite{Gadde:2011uv}, is defined by
\begin{align}
I_N=\tr[(-1)^F q^{J_1}x^{R_x}y^{R_y}],\quad(q=xy).
\label{schurdef}
\end{align}
See Appendix \ref{appB.sec} for the definition of Cartan generators
$H$, $J_1$, $J_2$, $R_x$, $R_y$, and $R_z$ of the
superconformal algebra $psu(2,2|4)$.
Only BPS operators saturating the bound
\begin{align}
H\geq J_1+R_x+R_y
\label{schbound0}
\end{align}
contribute to the index.
Refer to \cite{Bourdier:2015wda,Pan:2021mrw,Hatsuda:2022xdv} for analytic formulas of Schur indices.
The trace is taken over the Hilbert space of the gauge theory in $S^3$.
The line operator index is defined by the same equation,
but the trace is taken over the Hilbert space of the system in $S^3$ with line operators inserted
as external sources.
Let $I_N$ be the Schur index of $U(N)$ SYM without line insertion, and
let $I_{N,R}$ be the index with a pair of lines in a representation $R$
and its conjugate representation $\bar R$.
(Two representations may not be the conjugate representation to each other, but we focus on such a case.)
The expectation value is the ratio of these two indices:
\begin{align}
\langle W_R\bar W_R\rangle_N=\frac{I_{N,R}}{I_N}.
\end{align}
$W_R$ is the Wilson line operator in a representation $R$ and $\bar W_R=W_{\bar R}$.
On the gauge theory side, this quantity can be calculated
with the localization method,
and in the large $N$ limit, there exists a simple
analytic expression for it \cite{Hatsuda:2023imp}.
For the line operators in the fundamental representation $R={\rm fund}$,
it is given by \cite{Gang:2012yr}
\begin{align}
\langle W_{\rm fund}\bar W_{\rm fund}\rangle_\infty=I_{\rm F1}=\Pexp i_{\rm F1}=\frac{1-q}{(1-x)(1-y)},
\end{align}
where $\Pexp$ is the plethystic exponential and $i_{\rm F1}$ is the letter index
\begin{align}
i_{\rm F1}=x+y-q.
\label{if1}
\end{align}
This is reproduced as the index of fields on the worldsheet of the fundamental string \cite{Drukker:2000ep,Faraggi:2011bb}.

For a general irreducible representation
$R$ specified by a partition $\mu$, it is given by
\begin{align}
\langle W_R \bar W_R\rangle_\infty=
\sum_{\lambda\vdash|\mu|}\frac{1}{z_\lambda}|\chi_\mu(\lambda)|^2
[I_{\rm F1}]_\lambda.
\label{wgeneral}
\end{align}
$\sum_{\lambda\vdash|\mu|}$ denotes the sum over all partitions $\lambda$ of $|\mu|$.
$\chi_\mu(\lambda)$ is the character of a representation of the
symmetric group $S_k$ labeled by a partition $\mu$ evaluated at
the conjugacy class specified by a partition $\lambda$.
$z_\lambda$ is the integer
\begin{align}
z_\lambda=\prod_{m=1}^\infty m^{r_m}r_m!,
\end{align}
where $r_m$ is the number of occurrences of $m$ in $\lambda$.
$[\cdots]_\lambda$ is defined by
\begin{align}
[f(x)]_\lambda=\prod_{i=1}^{\ell(\lambda)}f(x^{\lambda_i}),
\end{align}
where $\ell(\lambda)$ is the length of $\lambda$.

For the totally anti-symmetric tensor representation $A_k$ of rank $k$,
the corresponding partition is $\mu=\{1^k\}$,
and the character for an arbitrary partition $\lambda$ of $k$ is $\chi_{\{1^k\}}(\lambda)=\pm1$.
The large $k$ limit of (\ref{wgeneral}) is \cite{Gang:2012yr}
\begin{align}
\lim_{k\rightarrow\infty}
\langle W_{A_k}\bar W_{A_k}\rangle_\infty
=I_{\rm D5}=\Pexp i_{\rm D5},\quad
i_{\rm D5}=\frac{1-q}{(1-x)(1-y)}-1.
\label{id5}
\end{align}
This is reproduced as the index of
the fluctuation modes on the corresponding D5-brane \cite{Faraggi:2011bb,Faraggi:2011ge}.
The form of the letter index suggests the structure of the brane system.
The two factors in the denominator, $1-x$ and $1-y$ correspond
to the tower of Kaluza-Klein modes carrying the charges $R_x$ and $R_y$,
which are Kaluza-Klein momenta in $S^4\subset S^5$.
Furthermore, it is also possible to reproduce the index with finite $k$
on the AdS side.
Just like the giant graviton expansions
\cite{Arai:2019xmp,Arai:2020qaj,Imamura:2021ytr,Gaiotto:2021xce,Murthy:2022ien}
for indices without operator insertions,
it is given in the form of the expansion \cite{Imamura:2024pgp}
\begin{align}
\langle W_{A_k}\bar W_{A_k}\rangle_\infty
=
I_{\rm D5}\sum_{m_x,m_y=0}^\infty x^{km_x}y^{km_y}I_{m_x,m_y},
\label{d5expansion}
\end{align}
where $I_{m_x,m_y}$ is the index of the theory realized on the system
of branes consisting of $m_x$ D3-branes on $D_x$ and $m_y$ D3-branes on $D_y$,
where $D_x$ and $D_y$ are three-dimensional disks in $S^5$ ending on the D5-brane worldvolume.
In the large $k$ limit, only $I_{0,0}=1$ contributes to the sum,
and (\ref{d5expansion}) reduces to (\ref{id5}).
We can also reproduce indices with both $N$ and $k$ being finite
by a similar expansion \cite{Imamura:2024pgp}.

Disappointingly, the success does not continue in the case of the symmetric representation $S_k$.\footnote{In the following we use $S_k$
to denote the symmetry group.}
With the character $\chi_{\{k\}}(\lambda)=1$
for the partition $\mu=\{k\}$ for the symmetric representation,
the formula (\ref{wgeneral}) gives the same result
as (\ref{d5expansion});
\begin{align}
\langle W_{S_k}\bar W_{S_k}\rangle_\infty=
\langle W_{A_k}\bar W_{A_k}\rangle_\infty.
\label{skandak}
\end{align}
On the other hand, the fluctuation modes on the D3-brane give
the index \cite{Faraggi:2011bb,Imamura:2024pgp}
\begin{align}
I_{\rm D3}=\Pexp i_{\rm D3},\quad
i_{\rm D3}=1-\frac{(1-x)(1-y)}{1-q}.
\label{id3}
\end{align}
The denominator $1-q$ in the letter index $i_{\rm D3}$ arises from the tower of Kaluza-Klein
modes in $S^2\subset AdS_5$.
This does not match the gauge theory result (\ref{skandak})
even in the large $k$ limit.\footnote{Another plausible candidate for the D3-brane line operator is the class of multiply wound Wilson lines in the fundamental representation,
which leads to a result distinct from that of the totally symmetric representation.
Unfortunately, it does not match the desired answer either.
See \cite{Hatsuda:2023iwi} for the detailed investigation of such operators on the gauge theory side, and \cite{Imamura:2024zvw} for its realization in terms of fundamental strings.}

Unfortunately, we have not found any solution to this problem
and will not discuss it further in this work.
One possibility is that the D3-brane corresponds to the insertion of
some other operators
labeled by $k$,
which we denote by $W_{{\rm D3},k}$.
Namely,
there may be an operator $W_{{\rm D3},k}$ such that
\begin{align}
\lim_{k\rightarrow\infty}\langle W_{{\rm D3},k}\bar W_{{\rm D3},k}\rangle_\infty=I_{\rm D3}.
\end{align}
Even though we do not understand the gauge theory description
of $W_{{\rm D3},k}$,
it is possible and important to study the D3-brane system on the AdS side in more detail.
A main purpose of this work is to study finite $k$ corrections to the D3-brane index.
As was pointed out in \cite{Imamura:2024pgp}, candidate objects contributing to the corrections
are
D1-branes stretched along a diameter of the $S^2$.
As we will explicitly show in Section \ref{finitek.sec},
they are BPS and contribute to the index.
The energy of a D1-brane is proportional to $k$, and
we expect the expansion
\begin{align}
\langle W_{{\rm D3},k}\bar W_{{\rm D3},k}\rangle_\infty=I_{\rm D3}\sum_{m=0}^\infty q^{mk}I_{{\rm D1},m},
\end{align}
where $m=0,1,2,\ldots$ is the number of D1-branes,
and $I_{{\rm D1},m}$ is the contribution from the modes on
$m$ coincident D1-branes stretched along the diameter of the $S^2$.
A main goal of this paper is to
determine the mode spectrum on a single D1-brane
and compute the corresponding letter index $i_{\rm D1}$,
which gives $I_{{\rm D1},1}$ in the leading finite $k$ correction by
\begin{align}
I_{{\rm D1},1}=\Pexp i_{\rm D1},
\label{id1pexp}
\end{align}
up to the zero-point contribution.

This paper is organized as follows.
After briefly reviewing the analysis of D3-brane fluctuations in the large $k$ limit in the next section,
we analyze fluctuation modes on a D1-brane.
In Section \ref{finitek.sec} we argue there must be boundary modes
localized near the endpoints of the D1-brane and guess the
spectrum using representation theory of the preserved symmetry.
The result is confirmed by a direct mode analysis in Section \ref{direct.sec}.
We will find the existence of modes belonging to non-unitary representations.
We will argue that they cause no problem and the Fock space still has
a positive norm in Section \ref{unitarity.sec}.
Section \ref{conc.sec} is devoted to conclusions and discussion.
Two appendices contain detailed explanations for conventions for spinors and indices.

\section{Large $k$ index from D3-brane}

The bosonic symmetry $so(2,4)_{\rm conf}\times so(6)_R\subset psu(2,2|4)$ of ${\cal N}=4$ SYM
is realized as the isometry of $AdS_5\times S^5$ on the gravity side.
To make it manifest,
it is convenient to define $AdS_5$ and $S^5$ as the subsets of the ambient spaces
$\RR^{2,4}$ and $\RR^6$:
\begin{align}
AdS_5 : \eta_{AB}X^AX^B=-1,\quad
S^5 : \delta_{KL}X^KX^L=1.
\end{align}
The indices $A$ and $B$ for $\RR^{2,4}$ run over six values $(\bullet,0,1,2,3,4)$,
and $K$ and $L$ for $\RR^6$ run over $(5,6,7,8,9,\circ)$.
The metric of $\RR^{2,4}$ is $\eta_{AB}=\diag(-,-,+,+,+,+)$.

The D3-brane configuration with $k$ units of the
electric flux on its worldvolume $AdS_2\times S^2\subset AdS_5$ is given by \cite{Drukker:2005kx}
\begin{align}
(X^\bullet)^2+(X^0)^2-(X^4)^2=1+\kappa^2,\quad
(X^1)^2+(X^2)^2+(X^3)^2=\kappa^2,
\label{d3wv}
\end{align}
and it is at $X^\circ=1$ in $S^5$.
$\kappa$ is the dimensionless parameter defined by
\begin{align}
\kappa=\frac{k}{2L^2T_{\rm D1}},
\label{kdef}
\end{align}
where $T_{\rm D1}$ is the D1-brane tension and $L$ is the AdS radius.
In the small $\kappa$ limit the D3-brane reduces to the
string worldsheet $AdS_2=AdS_5\cap\RR^{2,1}_{\bullet 04}$ for the fundamental representation.
(We use the notation $\RR^{2,1}_{\bullet 04}$ for the three-dimensional vector subspace
of the ambient space
along $\bullet 04$ directions.)

After the insertion of the D3-brane,
the unbroken symmetry is $osp(4^*|4)$, whose bosonic subalgebra is
$so(2,1)_{\bullet04}\times so(3)_{123}\times so(5)_{56789}\subset osp(4^*|4)$,
and its irreducible representations are specified by the
quantum numbers $H$, $J_1$ of the highest weights,
and an $so(5)_R=so(5)_{56789}$ representation $R_5$.
We use the notation $[J_1]_H^{R_5}$ for them.
It is shown in \cite{Faraggi:2011bb} that the
fluctuation modes on the D3-brane worldvolume belong to the $osp(4^*|4)$ representation
\begin{align}
{\cal R}_0
\oplus
{\cal R}_1
\oplus
{\cal R}_2
\oplus
{\cal R}_3
\oplus
\cdots
\end{align}
where each irreducible $osp(4^*|4)$ representation ${\cal R}_\ell$ ($\ell=0,1,2,\ldots$) consists of the
following components
\begin{align}
{\cal R}_0&=
[0]_{1}^{\bm5}
\oplus
[\tfrac{1}{2}]_{\frac{3}{2}}^{\bm{4}}
\oplus
[1]_{2}^{\bm{1}},\nonumber\\
{\cal R}_{\ell=1,2,\ldots}
&=[\ell-1]_{\ell}^{\bm1}
\oplus
[\ell-\tfrac{1}{2}]_{\ell+\frac{1}{2}}^{\bm{4}}
\oplus
[\ell]_{\ell+1}^{\bm5\oplus\bm{1}}
\oplus
[\ell+\tfrac{1}{2}]_{\ell+\frac{3}{2}}^{\bm{4}}
\oplus
[\ell+1]_{\ell+2}^{\bm{1}}.
\label{shortond3}
\end{align}
$\ell$ is the Kaluza-Klein momentum in the $S^2$ and
for the string worldsheet obtained in the small $S^2$ limit,
only ${\cal R}_0$ appears as the fluctuation modes
on the fundamental string worldsheet, and
the corresponding letter index is $i_{\rm F1}$ in (\ref{if1}).
Summing up all the contributions from the short representations in (\ref{shortond3}),
we obtain the letter index $i_{\rm D3}$ in (\ref{id3}).

\section{Finite $k$ correction}\label{finitek.sec}
\subsection{Classical contribution}

\begin{figure}[htb]
\centering
\includegraphics{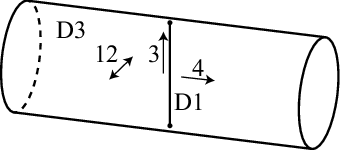}
\caption{The tubular D3-brane and a D1-brane along the diameter.}\label{dstring.eps}
\end{figure}

We consider a D1-brane
stretched along the diameter of the $S^2$ along the $X^3$ axis,
which is fixed under the action of $J_1$ (Figure \ref{dstring.eps}).
The worldsheet of the D1-brane (without excitation)
is the part of $AdS_2=AdS_5\cap\RR^{2,1}_{\bullet 03}$
restricted by $|X^3|\leq\kappa$.

Let us confirm the D1-brane carries quantum numbers saturating the BPS bound (\ref{schbound0}).
We parametrize the worldsheet by $(t,\sigma)$
and give the embedding of the D1-brane (without fluctuations) in $AdS_5$ by
\begin{align}
X^\bullet=\cosh\sigma\cos t,\quad
X^0=\cosh\sigma\sin t,\quad
X^3=\sinh\sigma,\quad
X^i=0\quad(i=1,2,4).
\label{tsigmadef}
\end{align}

The energy $H$, which is normalized by $1/L$ and dimensionless,
of the D-string without excitations is
\begin{align}
H=T_{\rm D1}L^2\int_{-\sigma_*}^{\sigma_*} d\sigma \cosh\sigma
=2T_{\rm D1}L^2\sinh\sigma_*
=k,
\end{align}
where $\sigma_*$ is defined by
\begin{align}
\kappa=\sinh\sigma_*.
\end{align}

In addition, the D-string possesses non-vanishing angular momentum $J_1$.
This comes from the coupling of the endpoints to the gauge field
on the D3-brane.
Let $\wt A$ be the dual gauge field minimally coupling to the
D1-brane endpoints with charge $\pm1$.
$\wt A$ for the $k$ unit of electric flux is
\begin{align}
\wt A=\frac{k}{2}\cos\theta d\phi,
\label{wta}
\end{align}
where $(\theta,\phi)$ are spherical coordinates on the $S^2$.
Let $\theta_\pm$ be the $\theta$ coordinates of two endpoints with charge $\pm1$.
We separate them into the values at the basepoints and fluctuations:
\begin{align}
\theta_+=0+\vartheta_+,\quad
\theta_-=\pi-\vartheta_-.
\end{align}
Then, the minimal coupling of the two endpoints
to (\ref{wta}) is given by the Lagrangian
\begin{align}
L=k\dot\phi
-\frac{k}{2}(1-\cos\vartheta_+)\dot\phi
-\frac{k}{2}(1+\cos\vartheta_-)\dot\phi.
\label{couplingtoaw}
\end{align}
The second and the third terms are contributions from the fluctuations.
From the first term $L^{(0)}=k\dot\phi$ we obtain
\begin{align}
J_1=\frac{\partial L^{(0)}}{\partial\dot\phi}=k.
\label{j1cls}
\end{align}

The D1-brane does not carry $R$ charges $R_x$ and $R_y$,
and these quantum numbers saturate the BPS bound (\ref{schbound0}).
The corresponding contribution to the index is $q^k$.

\subsection{Infinite D1-brane}

The D1-brane worldvolume with bosonic fluctuations
is given by the embedding
\begin{align}
X^\bullet&=\sqrt{1+\phi^2}\cosh\sigma\cos t,
&X^a&=\varphi_a\quad(a=56789),\nonumber\\
X^0&=\sqrt{1+\phi^2}\cosh\sigma\sin t,
&X^\circ&=\sqrt{1-\varphi^2},\nonumber\\
X^3&=\sqrt{1+\phi^2}\sinh\sigma,\nonumber\\
X^i&=\phi_i,\quad(i=124).
\label{d1flucads}
\end{align}
$\sigma$ and $t$ are coordinates on the worldsheet introduced in (\ref{tsigmadef}).
$\phi_i$ ($i=124$) and $\varphi_a$ ($a=56789$)
are scalar fields for fluctuations in
$AdS_5$ and $S^5$, respectively.
The action of bosonic fluctuations is obtained as the linearized part of the Nambu-Goto action
of the worldsheet defined by the embedding
(\ref{d1flucads})
\begin{align}
S_{\rm b1}=
T_{\rm D1}\int_{\rm D1}d\sigma dt\sqrt{-g}
\left(
-\tfrac{1}{2}(\partial_\mu\phi_i)^2
-\phi_i^2
-\tfrac{1}{2}(\partial_\mu\varphi_a)^2
\right).
\label{sb1}
\end{align}

We also have fermionic fields $\lambda$.
The quantum numbers of the fields are shown in Table \ref{d1fields.tbl}.
\begin{table}[htb]
\caption{Fields on the D1 brane}\label{d1fields.tbl}
\centering
\begin{tabular}{lccl}
\hline
\hline
fields & $so(3)_{124}$ & $so(5)_R$ \\
\hline
$\phi_i$ ($i=1,2,4$) & $\bm{3}$ & $\bm{1}$ & fluctuations in $AdS_5$ \\
$\varphi_a$ ($a=5,6,7,8,9$) & $\bm{1}$ & $\bm{5}$ & fluctuations in $S^5$ \\
$\lambda$ & $\bm{2}$ & $\bm{4}$ & fermions \\
\hline
\end{tabular}
\end{table}

The fluctuation modes of the fields on the infinite D-string without the restriction $|X^3|\leq\kappa$
are the same as those on the fundamental string corresponding to the
fundamental line operator
up to a certain change of the basis (the $3$-$4$ flip),
which does not affect the index.
They belong to the short representation ${\cal R}_0$ in (\ref{shortond3}),
and the letter index is the same as $i_{\rm F1}$ in (\ref{if1}).

\subsection{Introduction of the boundaries}
As we mentioned, the modes on an infinite string belong to the ${\cal R}_0$ representation
of $osp(4^*|4)$,
and the corresponding letter index is the same as
$i_{\rm F1}$ in (\ref{if1}).
In this section we introduce the boundaries at $\sigma=\pm\sigma_*$
and consider how the mode spectrum on the string is changed.
It does not agree with ${\cal R}_0$ even in the limit $\sigma_*\rightarrow\infty$
due to modes localized around the boundaries.
Let ${\cal R}_{\rm bdr}$ be the representation of modes localized around one of the boundaries.
Corresponding to the two boundaries related by a parity symmetry, two copies of ${\cal R}_{\rm bdr}$ appear,
and the large $\sigma_*$ limit of the spectrum on the segment is
\begin{align}
{\cal R}_0\oplus 2{\cal R}_{\rm bdr}.
\label{longlimit}
\end{align}
In the next section, we will determine the mode spectrum on
the segment by directly solving the wave equations for finite $\sigma_*$.
However,
we can guess the representation ${\cal R}_{\rm bdr}$ without direct calculations,
as we will show below,
and it is enough to calculate the index because the index should not depend on
the continuous parameter $\sigma_*$.

The superconformal symmetry realized on the infinite D-string
is $osp(4^*|4)$, and its bosonic subalgebra is
\begin{align}
so(2,1)_{\bullet03}\times so(3)_{124}\times so(5)_{56789}\subset osp(4^*|4).
\end{align}
The conformal generators $(H,P,K)$ of $so(2,1)_{\bullet03}$,
angular momenta generators $(J_1,J_\pm)$ of $so(3)_{124}$, $so(5)_R$ generators $R_{ab}$ ($a,b=5,6,7,8,9$),
and fermionic generators $(Q_{\pm,\alpha},S_\pm^\alpha)$ ($\alpha=1,2,3,4$)
carry the quantum numbers shown in Table \ref{generators.tbl}.
\begin{table}[htb]
\caption{Generators of $osp(4^*|4)$}\label{generators.tbl}
\centering
\begin{tabular}{c|ccc}
& $H$ & $J_1$ & $so(5)_R$ \\
\hline
$(H,P,K)$ & $(0,+1,-1)$ & $0$ & $\bm{1}$ \\
$(J_1,J_\pm)$ & $0$ & $(0,\pm1)$ & $\bm{1}$ \\
$R_{ab}$ & $0$ & $0$ & $\bm{10}$ \\
$(Q_{\pm,\alpha},S_\pm^\alpha)$ & $(+\frac{1}{2},-\frac{1}{2})$ & $\pm\frac{1}{2}$ & $\bm{4}$ \\
\end{tabular}
\end{table}
The introduction of the boundaries partially breaks the symmetry,
and only the commutant of $Z:=H+J_1$ is kept unbroken.
The unbroken subalgebra is $osp(2|4)\times so(2)_Z$,
where $so(2)_Z$ is generated by the central element $Z$,
and $osp(2|4)$ is generated by
\begin{align}
H-J_1,\quad
Q_\alpha\equiv Q_{-,\alpha},\quad
S^\alpha\equiv S_+^\alpha,\quad
R_{ab}.
\label{d1alg}
\end{align}
The unbroken supercharges satisfy
\begin{align}
\{Q_\alpha,S^\beta\}=\delta^\beta_\alpha(H-J_1)-\frac{i}{2}(\gamma^{ab})_\alpha{}^\beta R_{ab},
\label{qscomm}
\end{align}
where $\gamma^a$ are $so(5)_R$ Dirac matrices.

\begin{figure}[htb]
\centering
\includegraphics{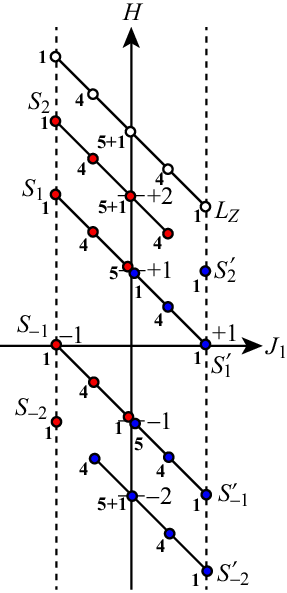}
\caption{Irreducible representations of $osp(2|4)$}\label{reps.eps}
\end{figure}

The modes on the infinite D-string belong to the representation ${\cal R}_0$ in (\ref{shortond3}),
which is decomposed into
the representations of the unbroken subalgebra labeled by $Z=1,2,3,\ldots$.\footnote{From the
purely algebraic point of view,
the structure of $osp(2|4)$ representations depends on the value of $H-J_1$ appearing in
(\ref{qscomm}) and has nothing to do with the value of $Z$.
However, in the irreducible representations appearing as the modes on the D1-brane,
$J_1$ is fixed by the quantum numbers of fields, and
$Z=H+J_1$ and $H-J_1$ are correlated.
(See Figure \ref{reps.eps}.)
For this reason we can label $osp(2|4)$ irreducible representations by $Z$.}
\begin{align}
{\cal R}_0=S_1\oplus
S_2\oplus
L_3\oplus
L_4\oplus
L_5\oplus\cdots.
\label{infinitestr}
\end{align}
$L_Z$ are long irreducible representations
(except for the special values of $Z$ which will be
shown below) with
the components
\begin{align}
L_Z=
[-1]_{Z+1}^{\bm1}\oplus
[-\tfrac{1}{2}]_{Z+\frac{1}{2}}^{\bm4}\oplus
[0]_Z^{\bm5\oplus\bm1}\oplus
[+\tfrac{1}{2}]_{Z-\frac{1}{2}}^{\bm4}\oplus
[+1]_{Z-1}^{\bm1}.
\end{align}
We use the notation $[J_1]_H^{R_5}$ for representations
of the unbroken bosonic symmetry.
$L_Z$ with $Z=\pm2,\pm1$ splits into two short irreducible representations:
\begin{align}
L_Z\rightarrow S_Z+S_Z',
\end{align}
where $S_Z$ and $S_Z'$ contain the following states:
\begin{align}
&S_2=
[-1]_3^{\bm1}\oplus
[-\tfrac{1}{2}]_{\frac{5}{2}}^{\bm4}\oplus
[0]_2^{\bm5\oplus\bm1}\oplus
[+\tfrac{1}{2}]_{\frac{3}{2}}^{\bm4},
\hspace{7em}
S_2'=[+1]_1^{\bm1},
\nonumber\\
&S_1=
[-1]_2^{\bm1}\oplus
[-\tfrac{1}{2}]_{\frac{3}{2}}^{\bm4}\oplus
[0]_1^{\bm5},
\hspace{5em}
S'_1=
[0]_1^{\bm1}\oplus
[+\tfrac{1}{2}]_{\frac{1}{2}}^{\bm4}\oplus
[+1]_0^{\bm1},
\nonumber\\
&S_{-1}=
[-1]_0^{\bm1}\oplus
[-\tfrac{1}{2}]_{-\frac{1}{2}}^{\bm4}\oplus
[0]_{-1}^{\bm1},
\hspace{2em}
S'_{-1}
=
[0]_{-1}^{\bm5}\oplus
[+\tfrac{1}{2}]_{-\frac{3}{2}}^{\bm4}\oplus
[+1]_{-2}^{\bm1}
\nonumber\\
&S_{-2}=[-1]_{-1}^{\bm1},
\hspace{4em}
S'_{-2}=
[-\tfrac{1}{2}]_{-\frac{3}{2}}^{\bm4}\oplus
[0]_{-2}^{\bm5\oplus\bm1}\oplus
[+\tfrac{1}{2}]_{-\frac{5}{2}}^{\bm4}\oplus
[+1]_{-3}^{\bm1}.
\end{align}
See also Figure \ref{reps.eps}.

Among infinite irreducible representations appearing in (\ref{infinitestr}),
only the two short multiplets $S_1$ and $S_2$ contribute to the Schur index.
Namely,
\begin{align}
i_{\rm F1}=i[S_1]+i[S_2].
\end{align}

Let us consider how
(\ref{longlimit}) is changed
when $\sigma_*$ decreases.
Let us first consider states in ${\cal R}_0$.
In particular, we focus on the modes of the $so(5)_R$ quintet scalar fields $\varphi_a$
appearing in every irreducible representation in (\ref{infinitestr}).
$\varphi_a$ are massless scalar fields in $AdS_2$ satisfying
$\Box\varphi_a=0$, where $\Box$ is the $AdS_2$ Laplacian.
The Dirichlet boundary condition is imposed at
the boundaries.
Therefore, the energy of each mode is a monotonically decreasing function
of $\sigma_*$.
Let $E_n(\sigma_*)$ ($n=1,2,3,\ldots$) be the energy of the $n$-th mode.
They have the following asymptotic form:
\begin{align}
E_n(\sigma_*)\stackrel{\sigma_*\rightarrow\infty}{\sim}n,\quad
E_n(\sigma_*)\stackrel{\sigma_*\rightarrow0}{\sim}\frac{\pi}{2\sigma_*}n.
\label{asymomega}
\end{align}
(An analytic expression for the energy eigenvalues
for an arbitrary $\sigma_*$ will be obtained in the next section.)
As $\sigma_*$ decreases,
the long representations $L_n$ ($n=3,4,5,\ldots$)
are continuously shifted to $L_{E_n(\sigma_*)}$.
This is also the case for the short representations $S_1$ and $S_2$,
which also contain the modes of $\varphi_a$ with
energy $E_k(\sigma_*)$ ($k=1,2$).
For this to be possible, 
they must be combined with
$S_1'$ and $S_2'$ to form long representations.
Therefore,
the boundary representation ${\cal R}_{\rm bdr}$
should contain these short representations, and
the minimum possibility for ${\cal R}_{\rm bdr}$ is
\begin{align}
{\cal R}_{\rm bdr}=S'_1\oplus S'_2.
\label{rbdr}
\end{align}
Indeed,
the modes in $S_1'$ carry the quantum numbers identical to
those of the broken generators $P$, $Q$, and $J_+$,
and the modes can be regarded as the Nambu-Goldstone modes
associated with the symmetry breaking due to the
boundaries.

If the minimality assumption
(\ref{rbdr})
is correct,
the spectrum for finite $\sigma_*$ becomes
\begin{align}
S_1'\oplus S_2'\oplus L_{E_1(\sigma_*)}\oplus L_{E_2(\sigma_*)}\oplus L_{E_3(\sigma_*)}\oplus\cdots
\label{segspec}
\end{align}
Because there are two copies of ${\cal R}_{\rm bdr}$,
one of ${\cal R}_{\rm bdr}$ is left out without being incorporated into the long representations
and contributes to the index.
Using the fact that long representations do not contribute to the index,
we obtain the letter index for the segment D-string as follows.
\begin{align}
i_{\rm D1}
=i[S_1']+i[S_2']
=-i[S_1]-i[S_2]
=-i_{\rm F1}=-x-y+q.
\label{naiveid1}
\end{align}
The corresponding multi-particle index is
\begin{align}
I_{{\rm D1},1}=\Pexp i_{\rm D1}=\frac{(1-x)(1-y)}{1-q}.
\label{naivei1}
\end{align}
This is a main result of this work.
Note that the zero-point contributions
from the three terms in (\ref{naiveid1}) cancel
and do not change (\ref{naivei1}).

Remark that $L_Z$ with $Z<2$ and $S'_1$ are not unitary representations.
Although they cause no inconsistency,
we need a special treatment, which modifies the single-particle spectrum (\ref{segspec}),
as we will discuss in Section \ref{unitarity.sec}.

\section{Direct mode analysis}\label{direct.sec}
\subsection{Supersymmetric action}
In this section we directly analyze the fluctuation modes, including fermionic ones.
We first define the local frame on the worldvolume
of the D1-brane using the section of the frame bundle (see Appendix \ref{appA.sec}.)
\begin{align}
g^{-1}=e^{i\sigma M^\bullet{}_3}e^{itM^\bullet{}_0}.
\end{align}
This corresponds to the embedding (\ref{tsigmadef})
with the vielbein
\begin{align}
e^t=\cosh\sigma dt,\quad
e^\sigma=d\sigma.
\end{align}

Let $\epsilon=(\epsilon_1,\epsilon_2)$ be the parameters for supersymmetry
transformations of type IIB supergravity.
It is an $so(2)_R$ doublet of $16$ component Majorana-Weyl spinors $\epsilon_i$ ($i=1,2$).
The general solution to the Killing spinor equation
(\ref{iibkilling}) is given
on the D1-brane worldvolume by
\begin{align}
\epsilon=\exp(\tfrac{\sigma}{2}\gamma^\bullet\gamma_3)
\exp(\tfrac{t}{2}\gamma^\bullet\gamma_0)\xi,
\label{d1killing}
\end{align}
where $\gamma^M$ are Dirac matrices acting on $so(2)_R$ doublet Majorana-Weyl spinors.
See Appendix \ref{appA.sec} for the definition.
$\xi$ is an arbitrary $so(2)$ doublet constant Majorana-Weyl spinor.
The condition for the supersymmetry preserved by the D1-brane is
\begin{align}
\gamma^{03}\sigma_x\epsilon=-\epsilon,
\label{d1susy}
\end{align}
and this is satisfied all over the D1-brane worldvolume
if
\begin{align}
\gamma^{03}\sigma_x\xi=-\xi.\label{d1susy0}
\end{align}

Using (\ref{d1susy}), we can rewrite (\ref{d1killing}) in the following $so(2)_R$ diagonal form.
\begin{align}
\epsilon=\exp(-\tfrac{\sigma}{2}\gamma^{1234}\sigma_z)
\exp(\tfrac{t}{2}\gamma^{0124}\sigma_z)\xi.
\label{d1killing2}
\end{align}

On the D1-brane worldvolume, $\epsilon_1$ and $\epsilon_2$ are
related by (\ref{d1susy}),
and we can use $\epsilon_1$ as $16$ independent parameters.
The Killing spinor equation satisfied by $\epsilon_1$
on the D1-brane worldvolume is
\begin{align}
D_\mu\epsilon_1=\tfrac{1}{2}\gamma^{124}\gamma_\mu\epsilon_1
\quad
(\mu=0,3)
\label{d1killing3}
\end{align}
The supersymmetric Lagrangian of the gauge multiplet on
the D1-brane is
\begin{align}
S_{\rm susy}=
T_{\rm D1}\int_{\rm D1}d\sigma dt\sqrt{-g}
\left(
-\tfrac{1}{2}(\partial_\mu\phi_i)^2
-\phi_i^2
-\tfrac{1}{2}(\partial_\mu\varphi_a)^2
-\tfrac{1}{2}(\lambda\sla D\lambda)
+\tfrac{1}{2}(\lambda\gamma^{124}\lambda)
\right).
\label{ld1}
\end{align}
(\ref{ld1}) is invariant under the supersymmetry transformations
\begin{align}
\delta\phi_i=(\lambda\gamma_i\epsilon_1),\quad
\delta\varphi_a=(\lambda\gamma_a\epsilon_1),\quad
\delta\lambda=(\sla\partial\sla\phi+\sla\partial\sla\varphi)\epsilon_1
-\gamma^{124}(\sla\phi+\sla\varphi)\epsilon_1,
\end{align}
where $\sla\phi=\phi_i\gamma^i$ and $\sla\varphi=\varphi_a\gamma^a$.

\subsection{Boundary conditions}
\paragraph{Scalar boundary conditions}
The D3-brane is a point in $S^5$, and the fields
$\varphi_a$ representing fluctuations in $S^5$ satisfy the Dirichlet boundary condition
\begin{align}
\varphi_a|_{\sigma=\pm\sigma_*}=0.
\label{varphibc}
\end{align}

The boundary conditions for $\phi_i$ ($i=124$) are neither Neumann nor Dirichlet because of the
bending of the D3-brane and the electric flux on the D3-brane.
Let us first consider the effect of the bending of the D3-brane.
Due to the bending, the value of the coordinate $\sigma$ at the endpoints may not be $\pm\sigma_*$.
Let $\sigma_\pm$ be the coordinates of the endpoints.
By substituting the embedding equations
(\ref{d1flucads}) into
the D3-brane worldvolume equation (\ref{d3wv})
we obtain
\begin{align}
\left((1+\phi_i^2)\sinh^2\sigma
+\phi_1^2
+\phi_2^2\right)\big|_{\sigma=\sigma_\pm}=\kappa^2.
\label{endsond3}
\end{align}
Namely, $\sigma_\pm$ depend on the
values of the scalar fields $\phi_i$ at the boundaries.
Let us define $\Delta\sigma_\pm$ as the difference of $\sigma_\pm$ from $\pm\sigma_*$ by
$\sigma_\pm=\pm(\sigma_*+\Delta\sigma_\pm)$.
(\ref{endsond3}) gives
\begin{align}
\Delta\sigma_\pm=
\left(-\frac{\cosh\sigma_*}{\sinh\sigma_*}\frac{\phi_1^2+\phi_2^2}{2}
-\frac{\sinh\sigma_*}{\cosh\sigma_*}\frac{\phi_4^2}{2}\right)\bigg|_{\sigma=\pm\sigma_*}.
\end{align}
$\Delta\sigma_\pm$ are negative if $\phi_i$ are non-vanishing at the ends of the D1-brane.
This changes the elastic energy of the D1-brane by $T_{\rm D1}\Delta\sigma_\pm$
and effectively induces the boundary potential terms
\begin{align}
S_{\rm b2}=T_{\rm D1}\cosh\sigma_*\sum_\pm\int dt
\left(\frac{\cosh\sigma_*}{\sinh\sigma_*}\frac{\phi_1^2+\phi_2^2}{2}
+\frac{\sinh\sigma_*}{\cosh\sigma_*}\frac{\phi_4^2}{2}\right)\bigg|_{\sigma=\pm\sigma_*},
\end{align}
where the summation is over the two boundary points.
The overall factor $\cosh\sigma_*$ comes from the metric.

The D1-brane endpoints carry Chan-Paton charges and minimally couple with the
dual potential field (\ref{wta}).
The second and the third terms in (\ref{couplingtoaw})
give the boundary action
\begin{align}
S_{\rm b3}
=\frac{T_{\rm D1}}{2\sinh\sigma_*}\sum_\pm
\int dt(-\phi_1\partial_t\phi_2+\phi_2\partial_t\phi_1)|_{\sigma=\pm\sigma_*}.
\end{align}
The variation of the bosonic action $S_{\rm b}=S_{\rm b1}+S_{\rm b2}+S_{\rm b3}$
gives the following boundary terms.
\begin{align}
\delta S_{\rm b}&=(\mbox{bulk terms})
\nonumber\\&
+T_{\rm D1}\sum_\pm\int dt
\bigg[
\pm\delta\phi_1
\left(-\cosh\sigma\partial^3\phi_1
+\frac{\cosh^2\sigma}{\sinh\sigma}\phi_1
-\frac{1}{\sinh\sigma}\partial_t\phi_2\right)
\nonumber\\&
\hspace{6em}
\pm\delta\phi_2
\left(-\cosh\sigma\partial^3\phi_2
+\frac{\cosh^2\sigma}{\sinh\sigma}\phi_2
+\frac{1}{\sinh\sigma}\partial_t\phi_1\right)
\nonumber\\&
\hspace{6em}
\pm\delta\phi_4
\left(-\cosh\sigma\partial^3\phi_4
+\sinh\sigma\phi_4\right)
\bigg]\bigg|_{\sigma=\pm\sigma_*}.
\end{align}
By requiring this to vanish for arbitrary $\delta\phi_i$,
the following boundary conditions are obtained.
\begin{align}
&
\left(
\sinh\sigma\partial_\sigma\phi_1-\cosh\sigma\phi_1+\frac{1}{\cosh\sigma}\partial_t\phi_2
\right)\bigg|_{\sigma=\pm\sigma_*}
=0,
\nonumber\\
&
\left(
\sinh\sigma\partial_\sigma\phi_2-\cosh\sigma\phi_2-\frac{1}{\cosh\sigma}\partial_t\phi_1
\right)\bigg|_{\sigma=\pm\sigma_*}
=0,
\nonumber\\
&\left(
\cosh\sigma\partial_\sigma\phi_4-\sinh\sigma\phi_4
\right)\big|_{\sigma=\pm\sigma_*}
=0.
\label{scalarbc0}
\end{align}

\paragraph{Fermion boundary conditions}
The presence of the D3-brane breaks supersymmetry in the same way as
the fundamental string does for the fundamental representation.
The preserved supersymmetry parameter satisfies
$\gamma^{04}\xi_1=\xi_1$,
and the corresponding Killing spinor $\epsilon_1$ satisfies
\begin{align}
(\gamma^{04}-e^{\sigma\gamma^{1234}})\epsilon_1=0,
\label{e1ond3}
\end{align}
and we can give $\epsilon_1$ satisfying this relation by
\begin{align}
\epsilon_1=(\gamma^{04}+e^{-\sigma\gamma^{1234}})\eta,
\end{align}
where $\eta$ is an arbitrary Majorana-Weyl spinor.
The fields $\varphi_a$ satisfy the Dirichlet boundary condition
$\varphi_a|_{\sigma=\pm\sigma_*}=0$ at the ends of the D1-brane, and then,
the supersymmetry transformation
\begin{align}
\delta\varphi_a
=(\lambda\gamma_a\epsilon_1)
=(\eta\gamma_a(\gamma^{04}-e^{-\sigma\gamma^{1234}})\lambda).
\end{align}
must also vanish at $\sigma=\pm\sigma_*$ for an arbitrary $\eta$.
This requires $\lambda$ to satisfy the boundary conditions
\begin{align}
(\gamma^{04}-e^{-\sigma\gamma^{1234}})\lambda|_{\sigma=\pm\sigma_*}=0.
\label{lambdabc}
\end{align}
As a consistency check, we can easily confirm that
the boundary conditions of scalar fields are reproduced from this condition.
Using (\ref{e1ond3}) we can rewrite
the supersymmetry transformation of (\ref{lambdabc}) as
\begin{align}
\left(\gamma^{04}-e^{-\sigma\gamma^{1234}}\right)\delta\lambda
&=2\gamma^{14}\epsilon_1\left(-\frac{1}{\cosh\sigma}\partial_t\phi_1+\sinh\sigma\partial_\sigma\phi_2-\cosh\sigma\phi_2\right)
\nonumber\\
&+2\gamma^{24}\epsilon_1\left(-\frac{1}{\cosh\sigma}\partial_t\phi_2-\sinh\sigma\partial_\sigma\phi_1+\cosh\sigma\phi_1\right)
\nonumber\\
&+2\gamma^{34}\epsilon_1\left(-\cosh\sigma\partial_\sigma\phi_4+\sinh\sigma\phi_4\right)
\nonumber\\
&-2e^{-\sigma\gamma^{1234}}\left(\partial_t\varphi_a\gamma^{a4}\epsilon_1
-\varphi_a\gamma^{124a}\epsilon_1\right).
\end{align}
For consistency, this must vanish at the endpoints $\sigma=\pm\sigma_*$,
and we obtain the boundary conditions
(\ref{varphibc}) and
(\ref{scalarbc0}).

\subsection{Fluctuation modes}
\paragraph{Fluctuations of $\varphi_a$}
The scalar fields
$\varphi_a$ ($a=56789$) are massless fields
representing
fluctuations in $S^5$ directions
and satisfy the wave equation $\Box\varphi_a=0$
where $\Box$ is the $AdS_2$ Laplacian.
Let us use the conformal gauge for the spatial coordinate on the D1-brane.
We introduce new coordinate $x$ by
\begin{align}
\sinh\sigma=\tan x,\quad
\cosh\sigma=\frac{1}{\cos x}.
\label{sigmatox}
\end{align}
(The first equation defines $x$, and the second equation is derived from the first.)
The Laplacian becomes $\Box=\cos^2x(-\partial_t^2+\partial_x^2)$.
Let $x_*$ be the value of $x$ corresponding to $\sigma_*$.

We take the ansatz
\begin{align}
\varphi_a(t,x)=e^{-i\omega t}f(x)
\end{align}
for each $a$.
The function $f(x)$ satisfies
the bulk equation
\begin{align}
(\partial_x^2+\omega^2)f=0,
\label{masslesskg}
\end{align}
and the boundary condition
\begin{align}
f(x)|_{x=\pm x_*}=0.
\label{dirichlet}
\end{align}
The solutions are
\begin{align}
f(x)=\sin\frac{\pi n(x_*-x)}{2x_*},\quad
\omega=n\omega_0,\quad
(n=\pm1,\pm2,\pm3,\ldots).
\label{fn123}
\end{align}
where $\omega_0$ is the parameter defined by
\begin{align}
\omega_0=\frac{\pi}{2x_*}.
\label{omega0}
\end{align}
The energy spectrum of these solutions with positive $n$ has the asymptotic behavior (\ref{asymomega}).

\paragraph{Fluctuations of $\phi_i$}

The scalar fields $\phi_i$ ($i=124$) represent
fluctuations in the $AdS_5$ directions.
They satisfy the
massive Klein-Gordon equation
\begin{align}
(\Box-2)\phi_i=0,
\label{kgmass2}
\end{align}
and the boundary conditions in (\ref{scalarbc0}).
We take the ansatz
\begin{align}
\phi_1(t,x)\pm i\phi_2(t,x)=e^{-i\omega t}f_\pm(x),\quad
\phi_4(t,x)=e^{-i\omega t}f_4(x).
\end{align}
The functions $f_4$ and $f_\pm$ satisfy
the bulk differential equation
\begin{align}
[(\partial_x+s)(\partial_x-s)+(\omega^2-1)]f_{4,\pm}=0,
\label{f4diffeq}
\end{align}
and the boundary conditions
\begin{align}
&[s(\partial_x-s)f_\pm- f_\pm\mp\omega f_\pm]|_{\rm bdr}=0
\nonumber\\
&[(\partial_x-s) f_4]|_{\rm bdr}=0,
\label{scalarbc2}
\end{align}
where $s$ is the function of $x$ defined by $s(x)=\sinh\sigma=\tan x$.
We use $(\cdots)|_{\rm bdr}$ instead of $(\cdots)|_{\sigma=\pm\sigma_*}$ to clarify
the plus-minus symbols in (\ref{scalarbc2}) are not correlated with $\pm\sigma_*$.

We can simplify equations by substituting
\begin{align}
f_{4,\pm}=(\partial_x+s)\hat f_{4,\pm}.
\label{ffromg}
\end{align}
We obtain the bulk equation
\begin{align}
(\partial_x^2+\omega^2)\hat f_{4,\pm}=0,
\label{gdiffeq}
\end{align}
and the boundary conditions
\begin{align}
&[(\partial_x\pm\omega s)\hat f_\pm]|_{\rm bdr}
=0,\nonumber\\
&\hat f_4|_{\rm bdr}=0.
\end{align}

If $\omega^2=1$, the relation
(\ref{ffromg}) is not invertible, and
there may be solutions that cannot be expressed in the form
(\ref{ffromg}).
We first discuss the generic case with $\omega^2\neq1$.
Then, we can invert (\ref{ffromg}) as
\begin{align}
\hat f_{4,\pm}=\frac{\partial_x-s}{1-\omega^2}f_{4,\pm}.
\end{align}
The exceptional case with $\omega^2=1$ will be discussed later separately.

Because the bulk equations and the boundary conditions are parity invariant,
we can consider even solutions and odd solutions separately,
and we do not have to consider their superpositions.
The even and odd solutions to the bulk equation
(\ref{gdiffeq}) are
\begin{align}
\hat f_{4,\pm}=\cos\omega x,\quad
\hat f_{4,\pm}=\sin\omega x,
\end{align}
respectively.
By imposing boundary conditions,
we obtain the following solutions for $\hat f_{4,\pm}$.
\begin{align}
\hat f_4&=\sin\omega(x-x_*),\quad\omega=n\omega_0,\quad (n=\pm1,\pm2,\pm3,\ldots),\nonumber\\
\hat f_\pm&=\cos(\omega(x-x_*)\pm x_*),\quad\omega=n\omega_0\pm1,\quad (n=\pm1,\pm2,\pm3,\ldots),\nonumber\\
\hat f_\pm&=1,\quad\omega=0.
\end{align}
From each of these solutions we obtain the corresponding $f_{4,\pm}$ by
(\ref{ffromg}).

For the exceptional values $\omega=\pm1$, we directly solve the original equations (\ref{f4diffeq}) and (\ref{scalarbc2}).
The even and odd solutions are
\begin{align}
f_{4,\pm}=\frac{1}{\cos x},\quad
f_{4,\pm}=\frac{x}{\cos x}+\sin x.
\end{align}
The first one satisfies the boundary condition (\ref{scalarbc2}) if $\omega=-1$ for $f_+$,
$\omega=+1$ for $f_-$, and $\omega=\pm1$ for $f_4$,
while the second one never satisfies the boundary condition.

\paragraph{Fluctuations of $\lambda$}

The fermion $\lambda$ satisfies the Dirac equation
\begin{align}
\sla D\lambda-\gamma^{124}\lambda=0,
\end{align}
and the boundary condition
(\ref{lambdabc}).
We adopt the following representation of the Dirac matrices
\begin{align}
\gamma^0&=i\sigma_x\otimes1_2\otimes 1_4\otimes \sigma_y,\nonumber\\
\gamma^1&=\sigma_y\otimes\sigma_x\otimes 1_4\otimes \sigma_y,\nonumber\\
\gamma^2&=\sigma_y\otimes\sigma_y\otimes 1_4\otimes \sigma_y,\nonumber\\
\gamma^3&=\sigma_z\otimes1_2\otimes 1_4\otimes \sigma_y,\nonumber\\
\gamma^4&=\sigma_y\otimes\sigma_z\otimes 1_4\otimes \sigma_y,\nonumber\\
\gamma^a&=1_2\otimes1_2\otimes\gamma^a_{(5)}\otimes \sigma_x
\quad (a=56789),\nonumber\\
\gamma^{11}&=1_2\otimes1_2\otimes1_4\otimes\sigma_z.
\end{align}
where $\gamma^a_{(5)}$ are $so(5)$ Dirac matrices satisfying $\gamma^{56789}_{(5)}=+1$.
Correspondingly, we take the ansatz
\begin{align}
\lambda=e^{-i\omega t}
\left(\begin{array}{c} f(\sigma) \\ g(\sigma) \end{array}\right)
\otimes\eta_{12}\otimes\eta_{56789}\otimes
\left(\begin{array}{c} 1 \\ 0 \end{array}\right),
\end{align}
where $\eta_{12}$ and $\eta_{56789}$ are
a $2$-component constant $so(2)_{12}$ spinor
and a $4$-component constant $so(5)_{56789}$ spinor,
respectively.

Using the vielbeins and the spin connection
\begin{align}
e^0=\cosh\sigma dt,\quad
e^3=d\sigma,\quad
\omega_{03}=-\tfrac{\sinh\sigma}{\cosh\sigma}e^0.
\end{align}
we can show that the Dirac equation and the boundary condition become
\begin{align}
\left(\begin{array}{cc}
\partial_\sigma+\tfrac{\sinh\sigma}{2\cosh\sigma} & \tfrac{\omega}{\cosh\sigma}-1 \\
-\tfrac{\omega}{\cosh\sigma}-1 & \partial_\sigma+\tfrac{\sinh\sigma}{2\cosh\sigma}
\end{array}\right)
\left(\begin{array}{c}
f \\ g
\end{array}\right)=0
\label{bulkeq1}
\end{align}
and
\begin{align}
\left(\begin{array}{cc}
\cosh\sigma+s_{12} & -\sinh\sigma \\
-\sinh\sigma & \cosh\sigma-s_{12}
\end{array}\right)
\left(\begin{array}{c}
f \\ g
\end{array}\right)\bigg|_{\sigma=\pm\sigma_*}=0.
\label{bdreq1}
\end{align}
where $s_{12}=\pm1$ is the eigenvalue of $2J_1=-i\gamma_{12}=\bm{1}\otimes\sigma_z\otimes\bm{1}_4\otimes\bm{1}_2$.
Namely, $\sigma_z\eta_{12}=s_{12}\eta_{12}$.

The bulk equation (\ref{bulkeq1}) and the boundary equation (\ref{bdreq1}) are
invariant under the replacement
\begin{align}
(f,g,\omega,s_{12})
\rightarrow (g,f,-\omega,-s_{12}).
\label{s12flip}
\end{align}
Let $\lambda_\pm$ be the components of $\lambda$ with
$s_{12}=\pm1$.
In the following, we derive the solutions for $\lambda_+$.
The solutions for $\lambda_-$ are obtained from them
by the replacement (\ref{s12flip}).

After the coordinate change (\ref{sigmatox}), the Dirac equation and the boundary conditions become
\begin{align}
\left(\begin{array}{cc}
\partial_x+\tfrac{\sin x}{2\cos x} & \omega-\tfrac{1}{\cos x} \\
-\omega-\tfrac{1}{\cos x} & \partial_x+\tfrac{\sin x}{2\cos x}
\end{array}\right)
\left(\begin{array}{c}
f \\ g
\end{array}\right)=0,\quad
\left(\frac{f}{g}-\tan\frac{x}{2}\right)\bigg|_{x=\pm x_*}=0.
\label{bulkbc0}
\end{align}

We can simplify these equations by substituting
\begin{align}
\left(\begin{array}{c}
f \\ g
\end{array}\right)
=
M\left(\begin{array}{c}
\hat f \\ \hat g
\end{array}\right),
\label{byfghat}
\end{align}
where $M$ is the matrix
\begin{align}
M=
\frac{1}{\sqrt{\cos x}}\left(\begin{array}{cc}
1-2\omega\cos x & \sin x \\
\sin x & 1+2\omega\cos x
\end{array}\right).
\end{align}
The determinant of the matrix is
\begin{align}
\det M=(1-4\omega^2)\cos x.
\end{align}
If $\omega=\pm\frac{1}{2}$ the matrix $M$ is singular, and there may be solutions
that cannot be expressed in the form 
(\ref{byfghat}).
We first consider the generic case with $\omega\neq\pm\frac{1}{2}$,
and the exceptional case with $\omega=\pm\frac{1}{2}$ will be discussed later.

After substitution of (\ref{byfghat}),
the bulk equation and the boundary conditions become
\begin{align}
\left(\begin{array}{cc}
\partial_x & -\omega \\
\omega & \partial_x
\end{array}\right)
\left(\begin{array}{c}
\hat f \\ \hat g
\end{array}\right)=0,\quad
\left(\frac{\hat f}{\hat g}+\tan\frac{x}{2}\right)\bigg|_{x=\pm x_*}=0.
\label{bulkbdr2}
\end{align}
The bulk equation has the following two linearly independent
solutions:
\begin{align}
\left(\begin{array}{c}
\hat f \\ \hat g
\end{array}\right)
=
\left(\begin{array}{c}
\sin\omega x \\
\cos\omega x
\end{array}\right),\quad
\left(\begin{array}{c}
\hat f \\ \hat g
\end{array}\right)
=
\left(\begin{array}{c}
\cos\omega x \\
-\sin\omega x
\end{array}\right).
\label{sols1}
\end{align}
Let us define the parity transformation
\begin{align}
{\cal P}:
(\hat f(x),\hat g(x))
\rightarrow (-\hat f(-x),\hat g(-x))
\end{align}
The first and the second solutions in 
(\ref{sols1}) are ${\cal P}$-even and ${\cal P}$-odd,
respectively.
We do not have to consider superpositions of (\ref{sols1})
because the bulk and the boundary equations in (\ref{bulkbdr2})
are invariant under ${\cal P}$.
The solutions satisfying the boundary conditions are
\begin{align}
\left(\begin{array}{c}
\hat f \\ \hat g
\end{array}\right)
=
\left(\begin{array}{c}
\sin(\omega(x-x_*)-\frac{x_*}{2}) \\
\cos(\omega(x-x_*)-\frac{x_*}{2})
\end{array}\right),\quad
\omega=n\omega_0-\frac{1}{2},\quad
n=\pm1,\pm2,\ldots.
\label{genericfmodes}
\end{align}
For even $n$ and odd $n$, this is ${\cal P}$-even and ${\cal P}$-odd, respectively.

Let us consider the exceptional cases with $\omega=\pm\frac{1}{2}$.
For $\omega=+\frac{1}{2}$,
The ${\cal P}$-even solution is
\begin{align}
\left(\begin{array}{c}
f \\ g
\end{array}\right)
=
\left(\begin{array}{c}
\frac{\sin\frac{x}{2}}{\sqrt{\cos x}}\\
\frac{\cos\frac{x}{2}}{\sqrt{\cos x}}
\end{array}\right).
\label{sol12a}
\end{align}
This satisfies the boundary condition in (\ref{bulkbc0})
regardless of the boundary position $x_*$.
The ${\cal P}$-odd solution
\begin{align}
\left(\begin{array}{c}
f \\ g
\end{array}\right)
=
\left(\begin{array}{c}
\sqrt{\cos x}\cos\frac{x}{2}
+\frac{x}{\sqrt{\cos x}}\sin\frac{x}{2}\\
\sqrt{\cos x}\sin\frac{x}{2}
+\frac{x}{\sqrt{\cos x}}\cos\frac{x}{2}
\end{array}\right)
\label{sol12b}
\end{align}
does not satisfy the boundary condition in (\ref{bulkbc0}) for $0<x_*<\frac{\pi}{2}$.
Two linearly independent solutions
for $\omega=-\frac{1}{2}$
are obtained from (\ref{sol12a}) and (\ref{sol12b}) by swapping $f$ and $g$,
and they never satisfy the boundary condition.

We summarize the results of the mode analysis in Table \ref{tb:bosonResult}.
We find the modes forming the representations in (\ref{segspec})
(and conjugate modes of them.)
\begin{table}[ht]
\caption{Energy eigenmodes on a segment D1-brane.
$n$ is an arbitrary non-zero integer.
This table includes both positive and negative frequency modes.}\label{tb:bosonResult}
\begin{center}
\begin{tabular}{c|cccc} \hline
 & $H$ & $J_1$ & $so(5)_R$ & rep. \\ \hline
$\phi_1+i\phi_2$ & $n\omega_0+1$ & $-1$ & $\bm{1}$ & $L_{n\omega_0}$ \\
                 & $0$           &      &          & $S_{-1}$ \\
                 & $-1$          &      &          & $S_{-2}$ \\
\hline
$\phi_1-i\phi_2$ & $n\omega_0-1$ & $+1$ & $\bm{1}$ & $L_{n\omega_0}$ \\
                 & $0$           &      &          & $S'_1$ \\
                 & $+1$          &      &          & $S'_2$ \\
\hline
$\phi_4$         & $n\omega_0$   & $0$  & $\bm{1}$ & $L_{n\omega_0}$ \\
                 & $+1$          &      &          & $S'_1$ \\
                 & $-1$          &      &          & $S_{-1}$ \\
\hline
$\varphi_a$      & $n\omega_0$   & $0$  & $\bm{5}$ & $L_{n\omega_0}$ \\
\hline
$\lambda_+$ & $n\omega_0-\frac{1}{2}$ & $+\frac{1}{2}$ & $\bm{4}$ & $L_{n\omega_0}$ \\
            & $+\frac{1}{2}$          &                &          & $S'_1$ \\
\hline
$\lambda_-$ & $n\omega_0+\frac{1}{2}$ & $-\frac{1}{2}$ & $\bm{4}$ & $L_{n\omega_0}$ \\
            & $-\frac{1}{2}$          &                &          & $S_{-1}$ \\
\hline
\end{tabular}
\end{center}
\end{table}

\section{Unitarity}\label{unitarity.sec}
As we mentioned in Section \ref{finitek.sec}, some boundary modes are below the BPS bound.
This does not mean any inconsistency.
Remember the existence of negative frequency modes
in the mode expansion of a field operator
does not cause any problem if we interpret the expansion coefficients
of the negative modes as creation operators.
Similarly, even if there exist modes below the BPS bound
and the corresponding state seems to have a negative norm,
we can construct positive norm a Fock space by appropriately
treating the corresponding operators.

Let $f_i$ and $g_k$ be bosonic and fermionic mode functions,
and $a_i$ and $b_k$ be the corresponding expansion coefficients.
Let us suppose $a_i$ and $b_k$ are annihilation operators,
and define the one-particle states by
\begin{align}
|f_i\rangle=a_i^\dagger|0\rangle,\quad
|g_k\rangle=b_k^\dagger|0\rangle.
\label{figk}
\end{align}
From these modes we obtain the letter index
\begin{align}
i=\sum_ix_i-\sum_kx_k,
\label{naiveletter}
\end{align}
where $x_i$ and $x_k$ are fugacities corresponding to $f_i$ and $g_k$, respectively.
Let us focus on a set of modes in an $osp(2|4)$ irreducible representation.
Single-particle states
$|f_i\rangle$ and
$|g_k\rangle$ in the representation
can be obtained from the primary state $|f_0\rangle$
by repeatedly applying raising operators $Q_\alpha$.
By using the algebra, we can determine the norms
$n_i=\langle f_i|f_i\rangle$ and
$n_k=\langle g_k|g_k\rangle$
of these states
up to a single overall factor determined by the norm of the primary state.
Then, the following (anti-)commutation relations hold:
\begin{align}
[a_i,a^\dagger_i]=n_i,\quad
\{b_k,b^\dagger_k\}=n_k.
\end{align}
If all $n_i$ and $n_k$ are positive,
all states in (\ref{figk}) are acceptable.
However, if some of the $n_i$ are negative,
we need to modify the definition of one-particle states.
Namely, for a bosonic mode with negative $n_i$,
we need to regard the corresponding operator $a_i$ as
a creation operator, and the corresponding one-particle
state is not one in (\ref{figk}) but
\begin{align}
|f_i^*\rangle=a_i|0\rangle.
\end{align}

This trick does not work for fermionic operators.
If $n_k$ is negative,
whether we regard $b_k$ as an annihilation operator or
as a creation operator,
the corresponding one-particle state becomes negative norm.
Therefore, the overall factor,
which we cannot determine by the algebra, should be chosen so that
all $n_k$ for fermionic modes are positive.
If this is impossible, we cannot avoid negative norm states.
(For an irreducible representation that does not contain fermionic modes,
the overall sign should be determined by an alternative criterion.)

Let us suppose the norms $n_k$ for the fermionic modes
are all positive, and we can form a positive norm Fock space.
The change of the interpretation of the expansion coefficients
$a_i$ affects the letter index (\ref{naiveletter}).
For a bosonic mode with negative $n_i$, we should replace
$x_i$ in the letter index
(\ref{naiveletter}) by $x_i^{-1}$.
This replacement
affects the multi-particle index only by the overall sign change.
For example, let us replace $x_i$ in the letter index by $x_i^{-1}$.
Before the replacement,
the contribution to the multi-particle index, including the zero-point factor, is
\begin{align}
x_i^{\frac{1}{2}}\Pexp( x_i)
=\frac{1}{x_i^{-\frac{1}{2}}-x_i^{\frac{1}{2}}}.
\label{mpxi}
\end{align}
By the replacement, this becomes
\begin{align}
x_i^{-\frac{1}{2}}\Pexp( x_i^{-1})
=\frac{1}{x_i^{\frac{1}{2}}-x_i^{-\frac{1}{2}}}.
\label{mpxi2}
\end{align}
(\ref{mpxi}) and 
(\ref{mpxi2}) differ by only the overall sign.
Therefore, $I_{{\rm D1},1}$ in (\ref{naivei1}) is still correct
up to the overall sign.

Let us determine the norms of (naively defined) one-particle states
in a long representation $L_Z$.
Let $s$ be the norm of the primary state $[+1]_{Z-1}^{\bm1}$.
Then, the other states in $L_Z$ have the following norms.
\begin{align}
[+1]_{Z-1}^{\bm1} &:s \nonumber\\
[+\tfrac{1}{2}]_{Z-\frac{1}{2}}^{\bm4} &:s(Z-2) \nonumber\\
[0]_{Z}^{\bm1} &:s(Z-2)(Z+1) \nonumber\\
[0]_{Z}^{\bm5} &:s(Z-2)(Z-1) \nonumber\\
[-\tfrac{1}{2}]_{Z+\frac{1}{2}}^{\bm4} &:s(Z-2)(Z-1)(Z+1) \nonumber\\
[-1]_{Z+1}^{\bm1} &:s(Z-2)(Z-1)(Z+1)(Z+2).
\label{norms}
\end{align}
What matters is only the sign of each norm in (\ref{norms}),
and its actual value is not important.
Except when $|Z|<1$, we can choose the sign of $s$ so that
all fermionic states have positive norms.
Therefore, representations with $|Z|>1$
are acceptable even if the naively constructed one-particle
states have negative norms.

For $Z=1$, the states in $L_1$ have the following norms.
\begin{align}
\begin{array}{cccccc}
[+1]_0^{\bm1} &
[+\tfrac{1}{2}]_{\frac{1}{2}}^{\bm4} &
[0]_1^{\bm1} &
[0]_1^{\bm5} &
[-\tfrac{1}{2}]_{\frac{3}{2}}^{\bm4} &
[-1]_2^{\bm1} \\
- & + & + & 0 & 0 & 0
\end{array}
\end{align}
By removing the zero-norm states, we obtain the representation
$S_1'$, and the null states form
the unitary representation $S_1$ by themselves.

For $Z=2$, the states in $L_2$ have the following norms.
\begin{align}
\begin{array}{cccccc}
[+1]_1^{\bm1} &
[+\tfrac{1}{2}]_{\frac{3}{2}}^{\bm4} &
[0]_2^{\bm1} &
[0]_2^{\bm5} &
[-\tfrac{1}{2}]_{\frac{5}{2}}^{\bm4} &
[-1]_3^{\bm1} \\
+s & 0 & 0 & 0 & 0 & 0
\end{array}
\end{align}
By removing the zero-norm states, we obtain the representation
$S_2'$, and the null states form
the unitary representation $S_2$ by themselves.

$S_2'$ does not contain fermionic modes,
and we cannot use the condition $n_k>0$ for fermionic modes
to determine the overall sign of the norm.
To determine whether we should treat the corresponding operator as
creation or annihilation, let us directly look at the
equation of motion and the time evolution by the Hamiltonian.
The mode in $S_2'$ is
\begin{align}
\phi^*=\frac{1}{\sqrt2}(\phi_1-i\phi_2)
=ce^{-it}\cosh\sigma=\frac{ce^{-it}}{\cos x}.
\label{s2mode}
\end{align}
The coefficient $c$ is the operator we want to determine whether it is annihilation or creation.
From the quantum numbers of the mode it must satisfy
\begin{align}
[H,c]=-c,\quad
[J_1,c]=-c.
\label{hcjc}
\end{align}
Let us calculate the Hamiltonian and
angular momentum $J_1$ for this mode.
The Lagrangian of the complex field $\phi=(\phi_1+i\phi_2)/\sqrt{2}$ including the
boundary term is
\begin{align}
L&=\int_{-\sigma_*}^{\sigma_*} d\sigma\cosh\sigma\left(\frac{1}{\cosh^2\sigma}|\partial_t\phi|^2
-|\partial_\sigma\phi|^2
-2|\phi|^2
\right)
\nonumber\\
&+\left[\frac{\cosh^2\sigma}{\sinh\sigma}|\phi|^2
-\frac{i}{2\sinh\sigma}(\phi\partial_t\phi^*-\phi^*\partial_t\phi)\right]^{\sigma_*}_{-\sigma_*}.
\end{align}
The Hamiltonian and the angular momentum $J_1$ are
\begin{align}
H&=\int_{-\sigma_*}^{\sigma_*} d\sigma\cosh \sigma\left(\frac{1}{\cosh^2\sigma}|\partial_t\phi|^2
+|\partial_\sigma\phi|^2
+2|\phi|^2
\right)
+\left[\frac{\cosh^2\sigma}{\sinh\sigma}|\phi|^2\right]^{\sigma_*}_{-\sigma_*},\nonumber\\
J_1&=\int_{-\sigma_*}^{\sigma_*} d\sigma\frac{i}{\cosh\sigma}(
\phi\partial_t\phi^*
-\phi^*\partial_t\phi
)-\left[\frac{1}{\sinh\sigma}|\phi|^2\right]_{-\sigma_*}^{\sigma_*}.
\end{align}
By substituting
(\ref{s2mode}),
we obtain
\begin{align}
H=J_1
=2|c|^2\left(\sinh\sigma_*-\frac{1}{\sinh\sigma_*}\right)
=-4|c|^2\cot(2x_*).
\label{hamofc}
\end{align}
The consistency of (\ref{hcjc}) and (\ref{hamofc}) requires
\begin{align}
[c^\dagger,c]=4\cot(2x_*).
\end{align}
If $x_*>\frac{\pi}{4}$, $c$ must be treated as an annihilation operator,
while if $x_*<\frac{\pi}{4}$, $c$ must be treated as a creation operator.

\begin{figure}[htb]
\centering
\includegraphics{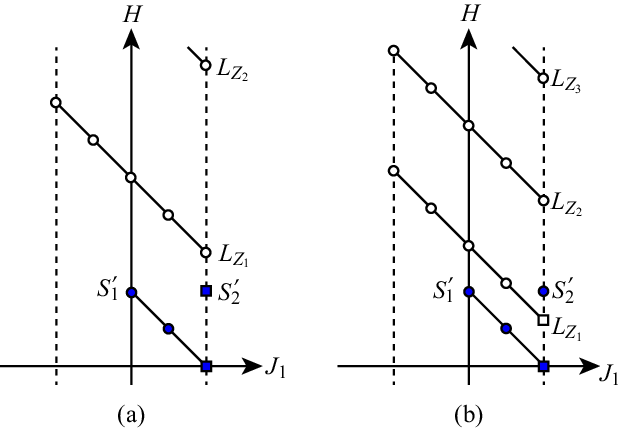}
\caption{The mode spectrum is shown in two cases:
(a) $\sinh\sigma_*<1$ and (b) $\sinh\sigma_*>1$.
For the modes shown as circles, we treat the corresponding expansion coefficients
in a standard way.
Namely, they are treated as annihilation operators.
For the modes shown as squares, the corresponding coefficients are treated as creation operators.}\label{spec.eps}
\end{figure}

The results are summarized in Figure \ref{spec.eps}.
In both cases with $x_*<\frac{\pi}{4}$ and $x_*<\frac{\pi}{4}$,
there are two bosonic states for which we need unusual treatment.
The naive contribution from each of them is $q$,
and it should be replaced by $q^{-1}$.
Namely, the letter index becomes
\begin{align}
i_{\rm D1}=i_{\rm D1}^{\rm naive}-2q+2q^{-1}.
\end{align}
This modification does not change the multi-particle index $I_{{\rm D1},1}$ in (\ref{naivei1}).


\section{Conclusions and discussion}\label{conc.sec}
In this work we considered AdS/CFT correspondence for ${\cal N}=4$ $U(N)$ SYM in the large $N$ limit,
and investigated line operators realized by D3-branes wrapped on $AdS_2\times S^2\subset AdS_5$.
It is labelled by a positive integer $k$, the number of the electric flux on the worldvolume.
In the large $k$ limit, we can calculate the corresponding line operator index by analyzing the fluctuation modes on the D3-brane.
The new result obtained in this work is the leading finite $k$ correction.
A D1-brane suspended along a diameter of the $S^2$ is BPS,
and it contributes to the index.
Starting from the result for an infinite D1-brane,
for which fluctuation modes had been known,
we argued that the introduction of the boundaries causes the emergence of boundary modes.
The consistency with representation theory
for the unbroken symmetry $osp(2|4)$ requires the existence of the boundary modes that break the unitarity bound.
We confirmed it by directly solving the wave equations
and boundary conditions for fields on the D1-brane.
Using the mode spectrum, we obtained the letter index (\ref{naiveid1})
and the leading correction (\ref{naivei1}) to the line-operator index.
We also discussed the presence of modes that belong to
non-unitary representations does not cause any problem
by treating the corresponding expansion coefficients in an appropriate manner.

We leave many questions for future works.
As is mentioned in Introduction, the most important question is
about the corresponding line operators on the gauge theory side.
In the literature, the D3-brane is usually regarded as the holographic dual of the
Wilson line operators in the totally symmetric representation of rank $k$.
However, the line-operator index for the symmetric lines
does not match the index of the line-operator index calculated with D3-brane.
It is desirable to resolve this issue before proceeding to
more detailed analysis of the D3-D1 system.

In this work, we consider only the leading finite $k$ corrections.
We expect higher-order corrections are obtained as contributions from
multiple D1-branes.
Unlike the leading contribution, which is the simple plethystic exponential
of the letter index and we can use the naively obtained letter index (\ref{naiveid1}),
we need to carefully choose the contours of gauge fugacity integrals.
In this process it may be important to use the correct spectrum
of physical excitations forming the positive-norm Fock space.

Another important problem is the finite $N$ corrections.
When $N$ is finite, just like the superconformal index
without line operator insertion,
D3-brane giants in $S^5$ can contribute to the index.
Furthermore, if there are D3-brane giants,
which locate at the center of $AdS_5$,
D1-branes giving the finite $k$ corrections can end on the D3-branes.
By analyzing such composite brane configurations,
it would be possible to obtain the line operator index for finite $N$ and finite $k$.

Furthermore, we can consider more general line operators realized by multiple D3-branes,
whose cross section is concentric multiple $S^2$.
In such a brane system,
not only D1-branes suspending along the diameter of each $S^2$,
but also BPS D1-branes connecting two $S^2$ would contribute to the index.

Investigation of these generalizations may be helpful
to establish the AdS/CFT dictionary for line operators.
We hope to return to these problems in the near future.

\section*{Acknowledgments}
The work of Y.~I. was supported by JSPS KAKENHI Grant Number JP21K03569,
and the work of A.~S was supported by JSPS KAKENHI Grant Number JP24KJ1105.

\appendix
\section{Killing spinor}\label{appA.sec}
Let us summarize the relation among
the $32$ supercharges, the corresponding transformation parameters $\xi$,
and the Killing spinors $\epsilon$.

We use the mostly plus metric.
Let $\epsilon=(\epsilon_1,\epsilon_2)$ be a pair of $16$-component Majorana Weyl spinors parameterizing the
local supersymmetry transformation of type IIB supergravity.
The chirality condition is
\begin{align}
\gamma^{11}\epsilon=+\epsilon,
\label{g11cond}
\end{align}
where the chirality matrix is defined by
$\gamma^{11}=\gamma^{012345678
9}$.
We use 01234 for $AdS_5$ and 56789 for $S^5$.

The symmetry algebra of the ${\cal N}=4$ SYM is
$psu(2,2|4)$.
In the classical theory we also have $so(2)_R$ symmetry,
which rotates $(\epsilon_1,\epsilon_2)$ as a doublet.

In addition to the ten-dimensional Dirac matrices $\gamma^M$ ($M=0,\ldots,9$),
we introduce
\begin{align}
\gamma^\bullet=\gamma^{56789}(i\sigma_y),\quad
\gamma^\circ=\gamma^{11}\gamma^\bullet,
\label{defbullet}
\end{align}
corresponding to the two extra coordinates $X^\bullet$ and $X^\circ$ in the ambient spaces.
$\sigma_i$ ($i=x,y,z$) are Pauli matrices acting on $so(2)_R$ indices.

Let $\epsilon$ be a Killing spinor satisfying the Killing spinor equation
\begin{align}
\delta\psi_M=(D_M-\tfrac{1}{2}\gamma^\bullet\gamma_M)\epsilon=0.
\label{iibkilling}
\end{align}
We can define the parameter $\xi$ for the rigid supersymmetry
as the value of the corresponding Killing spinor
$\epsilon$ evaluated at a reference point $P\in AdS_5\times S^5$:
\begin{align}
\xi=\epsilon|_P.
\label{xiandepsilon}
\end{align}
Because the ten differential operators
$D_M-\frac{1}{2}\gamma^\bullet\gamma_M$
appearing in (\ref{iibkilling})
all commute and
(\ref{iibkilling}) is integrable,
we can uniquely determine $\epsilon$ over the spacetime for a given $\xi$
by (\ref{iibkilling}) and (\ref{xiandepsilon}).
$\xi$ belongs to the bi-spinor representation of the bosonic subalgebra
$so(2,4)_{\rm conf}\times so(6)_R$.

Instead of directly solving the Killing spinor equation,
we can use isometry to generate the Killing spinor for a given $\xi$.
The fact that the Killing spinors belong to the bi-spinor representation of
$so(2,4)_{\rm conf}\times so(6)_R$ means that the Killing spinor
at $x\in AdS_5\times S^5$ is given by
\begin{align}
\epsilon(x)=\rho(g^{-1})\xi
\label{killingeq}
\end{align}
where $g\in SO(2,4)_{\rm conf}\times SO(6)_R$ is a rotation that takes the reference point $P$ to $x$,
and $\rho$ is the bi-spinor representation.
$g$ for each $x$ is not unique, and choosing $g(x)$ specifies the local frame at $x$.
(In other words, $g$ is a section of the frame bundle specifying the local frame.)

In this work we use the reference point $P$ with the coordinates
\begin{align}
P:(X^{\bullet},X^0,\ldots,X^4;X^5,\ldots,X^9,X^\circ)=(1,0,\ldots,0;0,\ldots,0,1),
\label{reference}
\end{align}
and assume $g=e$ at $x=P$.

The bi-spinor representation matrix $\rho$ is given explicitly by using the generating matrices
\begin{align}
iM^{AB}&=\tfrac{1}{2}\gamma^{AB}\quad(A,B=\bullet,0,1,2,3,4),\nonumber\\
iM^{KL}&=\tfrac{1}{2}\gamma^{KL}\quad(K,L=5,6,7,8,9,\circ).
\label{ads5s5rep}
\end{align}
$\gamma^A$ and $\gamma^K$
satisfy the $so(2,4)_{\rm conf}$ and $so(6)_R$ Clifford algebras, respectively,
and the matrices in (\ref{ads5s5rep}) satisfy the $so(2,4)\times so(6)$ algebra.
It is easy to confirm that
(\ref{killingeq}) satisfies the Killing spinor equation (\ref{iibkilling}).

\section{Superconformal and Schur indices}\label{appB.sec}
We define the six Cartan generators of $psu(2,2|4)$:
\begin{align}
H&=M_{\bullet 0},&
R_x&=M_{56},\nonumber\\
J_1&=M_{12},&
R_y&=M_{78},\nonumber\\
J_2&=M_{34},&
R_z&=M_{9\circ}.
\end{align}
We use the supercharge ${\cal Q}^\sci$ with the quantum numbers
\begin{align}
(H,J_1,J_2,R_x,R_y,R_z)
=(+\tfrac{1}{2},-\tfrac{1}{2},-\tfrac{1}{2},+\tfrac{1}{2},+\tfrac{1}{2},+\tfrac{1}{2}),
\label{qcharges}
\end{align}
to define the superconformal index.
\begin{align}
I=\tr[(-1)^Fq^{J_1}p^{J_2}x^{R_x}y^{R_y}z^{R_z}],\quad
qp=xyz.
\label{scidef}
\end{align}
The associated BPS bound is
\begin{align}
\{{\cal Q}^\sci,({\cal Q}^\sci)^\dagger\}
=H-J_1-J_2-R_x-R_y-R_z\geq0.
\label{scibound}
\end{align}
The reference point $P$ in (\ref{reference}) is
fixed under the actions of $J_1$, $J_2$, $R_x$, and $R_y$.
The Killing spinor $\epsilon^\sci$ corresponding to ${\cal Q}^\sci$
carries the quantum numbers opposite to (\ref{qcharges}),
and satisfies the following conditions:
\begin{align}
\gamma^{12}\xi^\sci&=+i\xi^\sci, & \gamma^{56}\xi^\sci&=-i\xi^\sci, & \gamma^{09}\xi^\sci&=+\xi^\sci,\nonumber\\
\gamma^{34}\xi^\sci&=+i\xi^\sci, & \gamma^{78}\xi^\sci&=-i\xi^\sci, & \sigma_y\xi^\sci&=-\xi^\sci.
\label{xisciqms}
\end{align}

We consider half-BPS line operator insertion in the boundary gauge theory.
The supersymmetry preserved by the line does not depend on the representation $R$.
The line operator of the fundamental representation
is realized by the worldsheet of a fundamental string
ending on the inserted lines on the
AdS boundary \cite{Rey:1998ik,Maldacena:1998im}.
We consider a fundamental string worldsheet on $AdS_2=AdS_5\cap\RR^{2,1}_{\bullet 04}$, which contains the reference point $P$.
It is half BPS, and
the condition for the supersymmetry preserved by the worldsheet is
\begin{align}
\gamma^{04}\sigma_z\xi=\xi.
\label{gamma042}
\end{align}
The superconformal index in (\ref{scidef})
is incompatible with the string insertion because
$\xi^\sci$ used to define the index does not satisfy (\ref{gamma042}).
Instead, we use the index associated with
the supercharge
${\cal Q}^\sch$
corresponding to the parameter
\begin{align}
\xi^\sch=\tfrac{1}{\sqrt2}(\xi^\sci+\gamma^{04}\sigma_z\xi^\sci)
\label{xisch}
\end{align}
satisfying the condition
(\ref{gamma042}).
The supercharge corresponding to $\gamma^{04}\sigma_z\xi^\sci$
in the second term in (\ref{xisch})
carries
the quantum numbers
\begin{align}
(H,J_1,J_2,R_x,R_y,R_z)
=(+\tfrac{1}{2},-\tfrac{1}{2},+\tfrac{1}{2},+\tfrac{1}{2},+\tfrac{1}{2},-\tfrac{1}{2}).
\end{align}
The BPS bound associated with ${\cal Q}^\sch$ is
\begin{align}
\{{\cal Q}^\sch,({\cal Q}^\sch)^\dagger\}=H-J_1-R_x-R_y\geq0,
\label{schbound}
\end{align}
and the index respecting ${\cal Q}^\sch$ is the Schur index
(\ref{schurdef}),
which is obtained from (\ref{scidef}) by taking the Schur limit $p=z$.
Note that the generators $J_1$, $R_x$, and $R_y$ appearing
in (\ref{schurdef})
are preserved by the string worldsheet.

Using (\ref{xisciqms}) we can show
$\gamma^{03}\sigma_x\xi^\sci
=-\gamma^{04}\sigma_z\xi^\sci$
and $\xi^\sch$ defined in (\ref{xisch}) satisfies
the condition (\ref{d1susy0}).
This means the D1-brane studied in the main text contributes to the Schur index.


\begin{thebibliography}{99}

\bibitem{Maldacena:1997re}
J.~M.~Maldacena,
``The Large N limit of superconformal field theories and supergravity,''
Adv. Theor. Math. Phys. \textbf{2}, 231-252 (1998)
doi:10.4310/ATMP.1998.v2.n2.a1
[arXiv:hep-th/9711200 [hep-th]].

\bibitem{Gubser:1998bc}
S.~S.~Gubser, I.~R.~Klebanov and A.~M.~Polyakov,
``Gauge theory correlators from noncritical string theory,''
Phys. Lett. B \textbf{428}, 105-114 (1998)
doi:10.1016/S0370-2693(98)00377-3
[arXiv:hep-th/9802109 [hep-th]].

\bibitem{Witten:1998qj}
E.~Witten,
``Anti-de Sitter space and holography,''
Adv. Theor. Math. Phys. \textbf{2}, 253-291 (1998)
doi:10.4310/ATMP.1998.v2.n2.a2
[arXiv:hep-th/9802150 [hep-th]].


\bibitem{Witten:1998zw}
E.~Witten,
``Anti-de Sitter space, thermal phase transition, and confinement in gauge theories,''
Adv. Theor. Math. Phys. \textbf{2}, 505-532 (1998)
doi:10.4310/ATMP.1998.v2.n3.a3
[arXiv:hep-th/9803131 [hep-th]].



\bibitem{Hosseini:2017mds}
S.~M.~Hosseini, K.~Hristov and A.~Zaffaroni,
``An extremization principle for the entropy of rotating BPS black holes in AdS$_{5}$,''
JHEP \textbf{07}, 106 (2017)
doi:10.1007/JHEP07(2017)106
[arXiv:1705.05383 [hep-th]].

\bibitem{Cabo-Bizet:2018ehj}
A.~Cabo-Bizet, D.~Cassani, D.~Martelli and S.~Murthy,
``Microscopic origin of the Bekenstein-Hawking entropy of supersymmetric AdS$_{5}$ black holes,''
JHEP \textbf{10}, 062 (2019)
doi:10.1007/JHEP10(2019)062
[arXiv:1810.11442 [hep-th]].

\bibitem{Choi:2018hmj}
S.~Choi, J.~Kim, S.~Kim and J.~Nahmgoong,
``Large AdS black holes from QFT,''
[arXiv:1810.12067 [hep-th]].

\bibitem{Amariti:2019mgp}
A.~Amariti, I.~Garozzo and G.~Lo Monaco,
``Entropy function from toric geometry,''
Nucl. Phys. B \textbf{973}, 115571 (2021)
doi:10.1016/j.nuclphysb.2021.115571
[arXiv:1904.10009 [hep-th]].



\bibitem{Chang:2013fba}
C.~M.~Chang and X.~Yin,
``1/16 BPS states in $\mathcal N=$ 4 super-Yang-Mills theory,''
Phys. Rev. D \textbf{88}, no.10, 106005 (2013)
doi:10.1103/PhysRevD.88.106005
[arXiv:1305.6314 [hep-th]].

\bibitem{Chang:2022mjp}
C.~M.~Chang and Y.~H.~Lin,
``Words to describe a black hole,''
JHEP \textbf{02}, 109 (2023)
doi:10.1007/JHEP02(2023)109
[arXiv:2209.06728 [hep-th]].

\bibitem{Choi:2022caq}
S.~Choi, S.~Kim, E.~Lee and J.~Park,
``The shape of non-graviton operators for SU(2),''
JHEP \textbf{09}, 029 (2024)
doi:10.1007/JHEP09(2024)029
[arXiv:2209.12696 [hep-th]].

\bibitem{Choi:2023znd}
S.~Choi, S.~Kim, E.~Lee, S.~Lee and J.~Park,
``Towards quantum black hole microstates,''
JHEP \textbf{11}, 175 (2023)
[erratum: JHEP \textbf{03}, 091 (2025)]
doi:10.1007/JHEP11(2023)175
[arXiv:2304.10155 [hep-th]].

\bibitem{Kim:2023sig}
S.~Kim, S.~Kundu, E.~Lee, J.~Lee, S.~Minwalla and C.~Patel,
``Grey Galaxies\textquoteright{} as an endpoint of the Kerr-AdS superradiant instability,''
JHEP \textbf{11}, 024 (2023)
doi:10.1007/JHEP11(2023)024
[arXiv:2305.08922 [hep-th]].


\bibitem{Choi:2023vdm}
J.~Choi, S.~Choi, S.~Kim, J.~Lee and S.~Lee,
``Finite N black hole cohomologies,''
JHEP \textbf{12}, 029 (2024)
doi:10.1007/JHEP12(2024)029
[arXiv:2312.16443 [hep-th]].

\bibitem{Choi:2024xnv}
S.~Choi, D.~Jain, S.~Kim, V.~Krishna, E.~Lee, S.~Minwalla and C.~Patel,
``Dual Dressed Black Holes as the end point of the Charged Superradiant instability in ${\cal N} = 4$ Yang Mills,''
[arXiv:2409.18178 [hep-th]].

\bibitem{deMelloKoch:2024pcs}
R.~de Mello Koch, M.~Kim, S.~Kim, J.~Lee and S.~Lee,
``Brane-fused black hole operators,''
[arXiv:2412.08695 [hep-th]].




\bibitem{Rey:1998ik}
S.~J.~Rey and J.~T.~Yee,
``Macroscopic strings as heavy quarks in large N gauge theory and anti-de Sitter supergravity,''
Eur. Phys. J. C \textbf{22}, 379-394 (2001)
doi:10.1007/s100520100799
[arXiv:hep-th/9803001 [hep-th]].

\bibitem{Maldacena:1998im}
J.~M.~Maldacena,
``Wilson loops in large N field theories,''
Phys. Rev. Lett. \textbf{80}, 4859-4862 (1998)
doi:10.1103/PhysRevLett.80.4859
[arXiv:hep-th/9803002 [hep-th]].

\bibitem{Drukker:2005kx}
N.~Drukker and B.~Fiol,
``All-genus calculation of Wilson loops using D-branes,''
JHEP \textbf{02}, 010 (2005)
doi:10.1088/1126-6708/2005/02/010
[arXiv:hep-th/0501109 [hep-th]].


\bibitem{Yamaguchi:2006tq}
S.~Yamaguchi,
``Wilson loops of anti-symmetric representation and D5-branes,''
JHEP \textbf{05}, 037 (2006)
doi:10.1088/1126-6708/2006/05/037
[arXiv:hep-th/0603208 [hep-th]].


\bibitem{Hartnoll:2006is}
S.~A.~Hartnoll and S.~P.~Kumar,
``Higher rank Wilson loops from a matrix model,''
JHEP \textbf{08}, 026 (2006)
doi:10.1088/1126-6708/2006/08/026
[arXiv:hep-th/0605027 [hep-th]].



\bibitem{Yamaguchi:2007ps}
S.~Yamaguchi,
``Semi-classical open string corrections and symmetric Wilson loops,''
JHEP \textbf{06}, 073 (2007)
doi:10.1088/1126-6708/2007/06/073
[arXiv:hep-th/0701052 [hep-th]].

\bibitem{Gomis:2006sb}
J.~Gomis and F.~Passerini,
``Holographic Wilson Loops,''
JHEP \textbf{08}, 074 (2006)
doi:10.1088/1126-6708/2006/08/074
[arXiv:hep-th/0604007 [hep-th]].

\bibitem{Gomis:2006im}
J.~Gomis and F.~Passerini,
JHEP \textbf{01}, 097 (2007)
doi:10.1088/1126-6708/2007/01/097
[arXiv:hep-th/0612022 [hep-th]].


\bibitem{Yamaguchi:2006te}
S.~Yamaguchi,
``Bubbling geometries for half BPS Wilson lines,''
Int. J. Mod. Phys. A \textbf{22}, 1353-1374 (2007)
doi:10.1142/S0217751X07035070
[arXiv:hep-th/0601089 [hep-th]].


\bibitem{Lunin:2006xr}
O.~Lunin,
``On gravitational description of Wilson lines,''
JHEP \textbf{06}, 026 (2006)
doi:10.1088/1126-6708/2006/06/026
[arXiv:hep-th/0604133 [hep-th]].

\bibitem{DHoker:2007mci}
E.~D'Hoker, J.~Estes and M.~Gutperle,
``Gravity duals of half-BPS Wilson loops,''
JHEP \textbf{06}, 063 (2007)
doi:10.1088/1126-6708/2007/06/063
[arXiv:0705.1004 [hep-th]].



\bibitem{Romelsberger:2005eg}
C.~Romelsberger,
``Counting chiral primaries in N = 1, d=4 superconformal field theories,''
Nucl. Phys. B \textbf{747}, 329-353 (2006)
doi:10.1016/j.nuclphysb.2006.03.037
[arXiv:hep-th/0510060 [hep-th]].

\bibitem{Kinney:2005ej}
J.~Kinney, J.~M.~Maldacena, S.~Minwalla and S.~Raju,
``An Index for 4 dimensional super conformal theories,''
Commun. Math. Phys. \textbf{275}, 209-254 (2007)
doi:10.1007/s00220-007-0258-7
[arXiv:hep-th/0510251 [hep-th]].




\bibitem{Dimofte:2011py}
T.~Dimofte, D.~Gaiotto and S.~Gukov,
``3-Manifolds and 3d Indices,''
Adv. Theor. Math. Phys. \textbf{17}, no.5, 975-1076 (2013)
doi:10.4310/ATMP.2013.v17.n5.a3
[arXiv:1112.5179 [hep-th]].

\bibitem{Gang:2012yr}
D.~Gang, E.~Koh and K.~Lee,
``Line Operator Index on $S^{1}\times S^{3}$,''
JHEP \textbf{05}, 007 (2012)
doi:10.1007/JHEP05(2012)007
[arXiv:1201.5539 [hep-th]].


\bibitem{Drukker:2015spa}
N.~Drukker,
``The $ \mathcal{N}=4 $ Schur index with Polyakov loops,''
JHEP \textbf{12}, 012 (2015)
doi:10.1007/JHEP12(2015)012
[arXiv:1510.02480 [hep-th]].



\bibitem{Hatsuda:2023iwi}
Y.~Hatsuda and T.~Okazaki,
``Exact $ \mathcal{N} $ = 2$^{*}$ Schur line defect correlators,''
JHEP \textbf{06}, 169 (2023)
doi:10.1007/JHEP06(2023)169
[arXiv:2303.14887 [hep-th]].

\bibitem{Guo:2023mkn}
Z.~Guo, Y.~Li, Y.~Pan and Y.~Wang,
``N=2 Schur index and line operators,''
Phys. Rev. D \textbf{108}, no.10, 106002 (2023)
doi:10.1103/PhysRevD.108.106002
[arXiv:2307.15650 [hep-th]].


\bibitem{Hatsuda:2023imp}
Y.~Hatsuda and T.~Okazaki,
``Large N and large representations of Schur line defect correlators,''
JHEP \textbf{01}, 096 (2024)
doi:10.1007/JHEP01(2024)096
[arXiv:2309.11712 [hep-th]].



\bibitem{Gadde:2011uv}
A.~Gadde, L.~Rastelli, S.~S.~Razamat and W.~Yan,
``Gauge Theories and Macdonald Polynomials,''
Commun. Math. Phys. \textbf{319}, 147-193 (2013)
doi:10.1007/s00220-012-1607-8
[arXiv:1110.3740 [hep-th]].


\bibitem{Bourdier:2015wda}
J.~Bourdier, N.~Drukker and J.~Felix,
``The exact Schur index of $\mathcal{N}=4$ SYM,''
JHEP \textbf{11}, 210 (2015)
doi:10.1007/JHEP11(2015)210
[arXiv:1507.08659 [hep-th]].


\bibitem{Pan:2021mrw}
Y.~Pan and W.~Peelaers,
``Exact Schur index in closed form,''
Phys. Rev. D \textbf{106}, no.4, 045017 (2022)
doi:10.1103/PhysRevD.106.045017
[arXiv:2112.09705 [hep-th]].


\bibitem{Hatsuda:2022xdv}
Y.~Hatsuda and T.~Okazaki,
``$ \mathcal{N} $ = 2$^{*}$ Schur indices,''
JHEP \textbf{01}, 029 (2023)
doi:10.1007/JHEP01(2023)029
[arXiv:2208.01426 [hep-th]].



\bibitem{Drukker:2000ep}
N.~Drukker, D.~J.~Gross and A.~A.~Tseytlin,
``Green-Schwarz string in AdS(5) x S**5: Semiclassical partition function,''
JHEP \textbf{04}, 021 (2000)
doi:10.1088/1126-6708/2000/04/021
[arXiv:hep-th/0001204 [hep-th]].


\bibitem{Faraggi:2011bb}
A.~Faraggi and L.~A.~Pando Zayas,
``The Spectrum of Excitations of Holographic Wilson Loops,''
JHEP \textbf{05}, 018 (2011)
doi:10.1007/JHEP05(2011)018
[arXiv:1101.5145 [hep-th]].

\bibitem{Faraggi:2011ge}
A.~Faraggi, W.~Mueck and L.~A.~Pando Zayas,
``One-loop Effective Action of the Holographic Antisymmetric Wilson Loop,''
Phys. Rev. D \textbf{85}, 106015 (2012)
doi:10.1103/PhysRevD.85.106015
[arXiv:1112.5028 [hep-th]].

\bibitem{Arai:2019xmp}
R.~Arai and Y.~Imamura,
``Finite $N$ Corrections to the Superconformal Index of S-fold Theories,''
PTEP \textbf{2019}, no.8, 083B04 (2019)
doi:10.1093/ptep/ptz088
[arXiv:1904.09776 [hep-th]].



\bibitem{Arai:2020qaj}
R.~Arai, S.~Fujiwara, Y.~Imamura and T.~Mori,
Phys. Rev. D \textbf{101}, no.8, 086017 (2020)
doi:10.1103/PhysRevD.101.086017
[arXiv:2001.11667 [hep-th]].





\bibitem{Imamura:2021ytr}
Y.~Imamura,
``Finite-N superconformal index via the AdS/CFT correspondence,''
PTEP \textbf{2021}, no.12, 123B05 (2021)
doi:10.1093/ptep/ptab141
[arXiv:2108.12090 [hep-th]].


\bibitem{Gaiotto:2021xce}
D.~Gaiotto and J.~H.~Lee,
``The giant graviton expansion,''
JHEP \textbf{08}, 025 (2024)
doi:10.1007/JHEP08(2024)025
[arXiv:2109.02545 [hep-th]].

\bibitem{Murthy:2022ien}
S.~Murthy,
``Unitary matrix models, free fermions, and the giant graviton expansion,''
Pure Appl. Math. Quart. \textbf{19}, no.1, 299-340 (2023)
doi:10.4310/PAMQ.2023.v19.n1.a12
[arXiv:2202.06897 [hep-th]].


\bibitem{Imamura:2024pgp}
Y.~Imamura and M.~Inoue,
``Brane expansions for anti-symmetric line operator index,''
JHEP \textbf{08}, 020 (2024)
doi:10.1007/JHEP08(2024)020
[arXiv:2404.08302 [hep-th]].

\bibitem{Imamura:2024zvw}
Y.~Imamura, A.~Sei and D.~Yokoyama,
``Giant graviton expansion for general Wilson line operator indices,''
JHEP \textbf{09}, 202 (2024)
doi:10.1007/JHEP09(2024)202
[arXiv:2406.19777 [hep-th]].


\end{thebibliography}
\end{document}